\documentclass[runningheads]{svmult}

\usepackage{makeidx}   
\usepackage{graphicx}  
\usepackage{subeqnar}  
\usepackage{multicol}  
\usepackage{physprbb}  
\makeindex             

\begin{document}
\title*{Star Clusters in Interacting Galaxies:\protect\newline The
Case of M51}
\toctitle{Star Clusters in Interacting Galaxies:
\protect\newline The Case of M51}
%
%
\titlerunning{Star Clusters in M51}
%
\author{Nate Bastian\inst{1}
\and Henny J.G.L.M. Lamers\inst{1}}
\authorrunning{N. Bastian and H.J.G.L.M. Lamers}
%
%
\institute{Astronomical Institute, Utrecht University, Princetonplein 5,
NL-3584 CC Utrecht, The Netherlands}

\maketitle              

\section{Introduction}
We present the results of an analysis of a population of stellar clusters
in the interacting galaxy, M51, using {\it HST-WFPC2}
observations.  The observations were
made in five broad band filters; f336w (U), f439w (B), f555w (V),
f675w (R), f814w (I), and transformed to the closest corresponding
Cousins-Johnson filter magnitude using the equations defined by
Holtzmann \cite{holtzmann}.  Only sources that were
found in every filter were analysed.  By comparing the spectral
energy distribution  
(SED) of each source with those those of evolutionary spectral
synthesis models \cite{bruzual}, 
we have derived an age, mass, and extinction for each source. The
comparison was done using
the {\it 3DEF method} \cite{bik}, that is based on a least $\chi^{2}$
test.  Only sources that were well-fit ($\chi_{\nu}^{2} \le 1.0$) and
resolved (fwhm $\ge 1.5$ pixels \cite{bastian}) were used in this
study.  533 sources pass these criteria, and will be referred to as
clusters for the remainder of the paper.  With this sample we are
able to study the  
cluster formation history, the cluster initial mass
function (cIMF), and search for evidence of cluster disruption. 

\section{Distributions of the Resolved Population}
\subsection{Mass, Extinction and Luminosity profiles.}
\begin{figure}[]
\begin{center}
\includegraphics[width=0.75\textwidth]{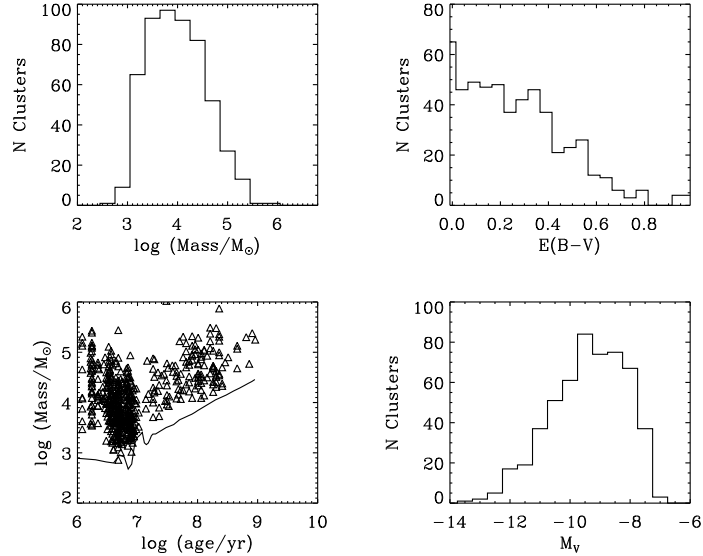}
\end{center}
\caption[]{The mass, extinction, age vs. mass, and absolute magnitude
 distributions of the 533 clusters that are well fit and resolved.}
\label{age-mass-mv}
\end{figure}

  In the following sections, all fits were done without the R band, as
  this band has been found to be heavily contaminated by H{$\alpha$}
emission.  Figure~\ref{age-mass-mv} shows the mass, extinction, and
  magnitude distributions of the cluster sample, as well as the age
  vs. mass diagram.  The age distribution will be discussed in
  Section~\ref{disruption}.  We see that our sample is incomplete
  fainter than 
  $M_{V}$ of -9.5 and cluster masses below $10^{4} M_{\odot}$.  The solid
  line in the age vs. mass distribution is the detection limit for
  $R_{lim} = 22.0$ mag.  As expected the lowest observable mass
  increases as a function of age (after 10 Myr), as the clusters fade
  due to stellar evolution.   

\subsection{Cluster IMF}
\begin{figure}[]
\begin{center}
\includegraphics[width=0.5\textwidth]{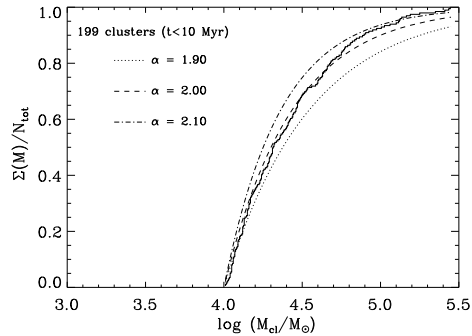}
\end{center}
\caption[]{The cumulative mass distribution of all well-fit clusters
  with ages $\le$ 10 Myr.  Over plotted are the power laws of the form
  $N(M_cl) \propto M_{cl}^{-\alpha}$ for $\alpha = 2.1, 2.0, 1.9$
  from top to bottom respectively.}
\label{imf-plot}
\end{figure}

In order to look at the cluster IMF, one needs to look at the
cluster population that has not been affected by disruption.  Recently,
Boutloukos \& Lamers (2002) have found that the characteristic disruption
timescale of a $10^{4} M_{\odot}$ cluster is $\sim 40 Myr$ in
M51, at a similar distance from the nucleus as the present
observations.  Due to the uncertainties in age fitting, we adopt a
conservative age cut-off of 10 Myr, when looking at the cluster IMF.
Figure~\ref{imf-plot} shows the cumulative mass distribution of all
well-fit clusters with ages $\le$ 10 Myr.  Clearly $\alpha = 2.0$ is
the best fit, though there is a significant deviation from this at
higher masses.  This may be evidence of a double power-law cIMF, though it can not be ruled out as an effect of uncertainties in the mass calculations.

\section{Evidence for Disruption}
\label{disruption}

Following the work of Boutloukos \& Lamers (2002, BL02), we have searched
our sample for evidence of cluster disruption.  Our sample has the
benefit over previous studies of disruption in M51 of covering a much
larger area of the galaxy.  The deficiency of this study is the
brighter detection limit which hampers our analysis.  We adopt the
same method to find the characteristic cluster disruption timescale as
BL02, which assumes that clusters disrupt as a function of their
initial mass as 
\begin{equation}
t_{dis}(M_{cl}) = t^{dis}_{4}(M_{cl}/10^{4}M_{\odot})^{\gamma},
\end{equation}
where $t^{dis}_{4}$ is the characteristic disruption timescale for a
$10^{4} M_{\odot}$ cluster.

\begin{figure}[]
\begin{center}
\includegraphics[width=0.85\textwidth]{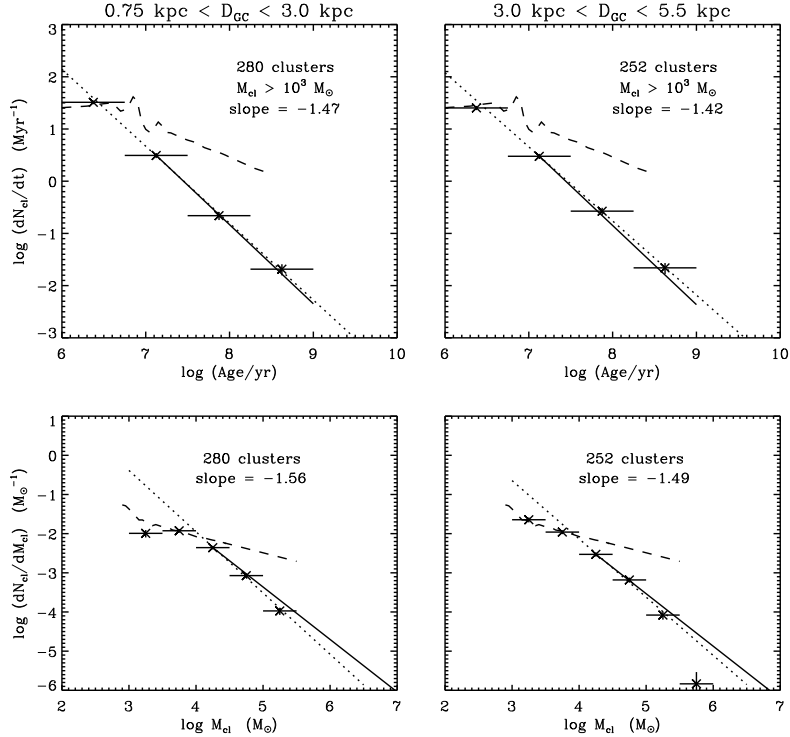}
\end{center}
\caption[]{The age (top figures) and mass (bottom figures) distributions for clusters broken into two
  different 
  distance bins (the left plots are for clusters from $0.75 < D_{GC}$
  kpc $< 3.0$, and the right plots are for clusters from $3.0 < D_{GC}$ kpc
 $< 5.5$, where $D_{GC}$ is the distance from the galactic center).  The dashed lines
  represent the expected fading line for a continuous formation
  history of clusters (note that the vertical scaling of this line is
  arbitrary and it is used here just to denote the expected
  distribution if disruption was not important). The solid line is the
  slope predicted by BL02 if $\gamma = 0.62$.  The dotted lines are
  the best fit to the data for clusters with $t_{cl} > 10^{7} yr$ and
  $M_{cl} > 10^{4} M_{\odot}$ (i.e. above the detection limit) for the
  age and mass distributions respectively.  The slope of this line is given for each distribution.}
\label{dndall}
\end{figure}

Fig.~\ref{dndall} shows the age (clusters formed per Myr) and mass
(clusters formed per $M_{\odot}$) distributions.  We see that the distributions show little to no
dependence on the distance from the center of M51, though we note that
the expected break in the mass distribution is below our detection
limit.  The last mass bin in the $3.0 < D_{GC} ({\rm kpc}) < 5.5$ sample has
only one cluster in it, and as such it should not be considered as a reliable data
point.  The age distributions do
not show any significant evidence for a burst in cluster formation
during the last Gyr.  This may be physical or it may be caused by a
low age resolution that we are able to reach with our data. The {\it
  3DEF} method takes into account the observational errors in
calculating the error associated with the determined age, and therefore
smaller observational errors result in a more refined age
determination.

Following the method prescribed in BL02, we have derived values for
$\gamma$, though we are not able to find values for $t_{dis}^{4}$ due
to the lack of an observed break in the age and mass distributions.
From the age distribution we derive $\gamma = 0.69 \pm 0.15$ in
excellent agreement with the findings of BL02, while
from the mass distributions we derive $\gamma = 0.47 \pm 0.15$
(assuming $\alpha = 2.0$). These two values suggest that $\gamma
\simeq 0.57 \pm 0.11$, in agreement with the mean value of $0.62 \pm 0.06$
found by BL02 in four different galaxies.

\section{Conclusions and Future Work}

\begin{itemize}
\item The cluster Initial Mass Function is well represented as a power
  law with an index of $-2.05 \pm 0.05$, although there is a significant
  deviation at higher masses

\item The derived values of the disruption parameter 
$\gamma$ and the exponent of the cluster IMF $\alpha$ is independent of 
the distance from the center of M51.

\item The data was not deep enough to trace the characteristic
  disruption timescale as a function of increasing distance from the
  galactic center.

\item We find values of $\gamma$ of $0.69 \pm 0.15$ and $0.47 \pm
  0.15$ from the age and mass distributions respectively.  These
  values are only marginally consistent with each other. 

\item We will look for the dependence of the cluster disruption
  timescale on the cluster radius,
  which will give a more precise description of the disruption,
  and can be compared with predictions.

\end{itemize}

%

\end{document}